\pdfoutput=1
\documentclass[aps,twocolumn,superscriptaddress,nofootinbib]{revtex4-2}
\usepackage{graphicx} 
\usepackage{amsthm} 
\usepackage{amsmath,braket}
\usepackage{mathtools}
\usepackage{bbold}  
\usepackage{amsfonts} 
\usepackage{physics} 
\usepackage{graphicx}
\usepackage[dvipsnames]{xcolor} 
\usepackage{tikz}
\usepackage{tikz-cd} 
\usepackage{adjustbox} 
\usepackage{centernot}
\usepackage[T1]{fontenc}
\usepackage[normalem]{ulem} 

\usetikzlibrary{decorations.pathreplacing}
\usetikzlibrary{decorations.pathmorphing}

\DeclareFontFamily{U}{mathb}{\hyphenchar\font45} 
\DeclareFontShape{U}{mathb}{m}{n}{
<-6> mathb5 <6-7> mathb6 <7-8> mathb7
<8-9> mathb8 <9-10> mathb9
<10-12> mathb10 <12-> mathb12
}{}
\DeclareSymbolFont{mathb}{U}{mathb}{m}{n}
\DeclareMathSymbol{\llcurly}{\mathrel}{mathb}{"CE}
\DeclareMathSymbol{\ggcurly}{\mathrel}{mathb}{"CF}

\newcommand{\thermomaj}{\succ_{\rm th}}

\theoremstyle{definition}

\begin{document}


\title{Quantum resource-theoretical analysis of the role\\ of vibrational structure in photoisomerization}

\author{Siddharth Tiwary}
\email{siddharth110200@gmail.com}
\affiliation{Department of Physics, Indian Institute of Technology Bombay, Powai, Mumbai 400076, India}
\affiliation{Department of Physics, 366 Le Conte \#7300, University of California, Berkeley, CA 94720, USA}
\author{Giovanni Spaventa}
\email{giovanni.spaventa@uni-ulm.de}
\author{Susana F. Huelga}
\email{susana.huelga@uni-ulm.de}
\author{Martin B. Plenio}
\email{martin.plenio@uni-ulm.de}
\affiliation{Institute of Theoretical Physics \& IQST, Ulm University, Albert-Einstein-Allee 11 89081, Ulm, Germany}

\begin{abstract}
Thermodynamical systems at the nanoscale, such as single molecules interacting with highly structured vibrational 
environments, typically undergo non-equilibrium physical processes that lack precise microscopic descriptions. 
Photoisomerization is such an example which has emerged as a platform on which to study single-molecule ultrafast 
photochemical processes from a quantum resource theoretic perspective. However, upper bounds on its efficiency 
have only been obtained under significant simplifications that make the mathematics of the resource-theoretical 
treatment manageable.
Here we generalize previous models for the photoisomers, while retaining the full vibrational structure, 
and still get analytical bounds on the efficiency of photoisomerization. We quantify the impact of such vibrational 
structure on the optimal photoisomerization quantum yield both when the vibrational coordinate has no dynamics of its own and when we take into account the
vibrational dynamics. This work serves as an example of how to bridge the gap between the abstract language of quantum
resource theories and the open system formulation of nanoscale processes.
\end{abstract}

\maketitle

{\emph{Introduction --}} Quantum resource theories (QRTs) \cite{coecke2016mathematical,chitambar2019quantum}, and in particular the formulations of thermodynamics that have emerged from them \cite{ruch1976principle,ruch1978mixing,janzing2000thermodynamic,horodecki2013fundamental,goold2016role,lostaglio2019introductory,ng2018resource}, represent a promising toolbox for the study of fundamental limitations to nanoscale processes \cite{halpern2020fundamental,spaventa2022capacity,burkhard2023boosting}, that are otherwise difficult to capture by solely relying on classical, equilibrium and macroscopic frameworks. In particular, QRTs offer the possibility of studying a physical process without making too many assumptions on the underlying microscopic theory that governs it, e.g. the specific structure of the environment, or the coupling strengths etc. This is possible because the theory only requires very general assumptions on the class of environments that are considered possible, for example on the grounds of symmetry considerations. However, applying the framework of QRTs of thermodynamics to realistic physical systems is a difficult task to say the least. Typically, models describing the systems of interest must be simplified considerably, before reaching a form that can easily be handled by the oftentimes convoluted state conversion criteria of resource theories. A relevant example is that of photoisomerization, a photochemical process that is at the basis of human vision \cite{schulten1978biomagnetic}, plays a key role in the primary steps of photosynthesis in plants, algae and bacteria \cite{croce2018light}, and can also be artificially controlled for technological applications, such as the storage of solar energy, nanorobotics and optical data storage \cite{dattler2019design}. Crucially, the microscopic details of the physics of photoisomerization is rather difficult to capture due to its non-equilibrium nature, its ultra-fast speed, the involvement of vibrational modes and its very high quantum yield \cite{nogly2018retinal,seidner1994microscopic,seidner1995nonperturbative,hahn2000quantum,hahn2002ultrafast}. For this reason, recent works \cite{halpern2020fundamental,spaventa2022capacity,burkhard2023boosting} have deployed the resource theory of athermality to find fundamental limitations to the efficiency of photoisomerization, in a way that is independent of the microscopic details governing the dynamics. However, the resource-theoretical model that has been used so far relies on very strong simplifying assumptions, that are needed in order to make efficient use of the thermomajorization conditions underpinning the state conversion criterion under \textit{thermal operations}. In particular, the photoisomer's vibrational structure is simplified, and reduced to the switching between the two stable configurations only, so as to keep the Hilbert space dimensionality small enough ($\leq 4$). Although this approximation might seem too heavy to be justified, its impact on the final result might be thought as somewhat limited by the fact that the only photoisomer states that are stable enough to have any practical functionality (and are therefore not merely transient states) must be sufficiently localised around the two stable configurations of the molecule, and as such can be approximated as having no support outside this reduced four-dimensional Hilbert space. However, one should consider that Hilbert space dimensionality can on its own play the role of a thermodynamical resource in physical processes \cite{silva2016performance}, and thus we can expect severe dimensionality cutoffs to result in an underestimation of the optimal efficiency of photoisomerization. The availability of multiple intermediate or final configurations may lead to an \textit{entropic} boost in the efficiency of photoisomerization. 
This motivates the next step of relaxing this simplifying assumption with a two-fold goal in mind: one, to strengthen the previous results by generalising the model; the other, to show that the framework of QRTs can be flexible enough to accomodate for more realistic and complex descriptions of nanoscale processes without giving up their predictive power.

{\emph{Outline --}} The paper is organised as follows. The first section gives a short introduction to the resource theory of athermality. Then, we proceed to describe how to build a resource-theoretical model for photoisomerization that does not rely on the previously needed simplifying assumptions. Finally, we present the main results in two sections: the first one treats the vibrational degree of freedom as a passive classical label, while the second takes into account the vibrational kinetic contribution and quantifies its impact on the photoisomerization yield. Finally, the results are compared and discussed.

{\emph{Thermodynamics as a quantum resource theory --}} The framework of quantum resource theories (QRTs) allows the rigorous quantification of physical resources (such as quantum coherence, entanglement, non-Markovianity etc.) and their interconversion. They provide a theoretical framework in which a set of operations (i.e. a subset of all quantum channels) are considered \textit{free}, and any state that cannot be prepared via free operations is then singled out as a \textit{(static) resource}, in the sense of facilitating a task inaccessible to the free set. Non-free states (or operations) can thus only be prepared (or implemented) at a cost, while on the other hand assisting processes that would be otherwise impossible or only attainable with a smaller fidelity. A prominent example of a resource theory is the theory of bipartite entanglement \cite{plenio2007introduction,horodecki2009quantum}, where the restriction to local operations and classical communication (LOCC) singles out entanglement as a resource, while separable states are considered free. Crucially, quantum thermodynamics can also be formulated as a resource theory, in which the Gibbs state is the only free state \cite{lostaglio2019introductory,ng2018resource}. As for the allowed operations, different choices are possible depending on the specific physical scenario, and lot of progress has been made in studying the thermodynamical properties of these sets of operations and the relationship between them \cite{de2024entanglement,son2024catalysis,ding2021exploring,vom2022bath,vom2023exploring}. In this work, we focus on the resource theory of \emph{athermality}, whose operations, called \textit{thermal operations}, are constructed as follows. Given a system $S$ with Hamiltonian $H_S$, the following three elementary operations are allowed: (i) The system can be brought into contact with a thermal bath $B$, that is, we can freely deploy Gibbs states $\tau = e^{-\beta H_B}/Z$ at inverse temperature $\beta$. (ii) We can perform any global unitary transformation $U$ on $S+B$, as long as it is energy preserving, i.e., $ \big[ U,H_S+H_B \big]=0$ . (iii) We are allowed to trace out subsystems, and in particular the entire bath $B$.
As a result, the action of \emph{thermal operations} (TO) on a density operator $\rho_S$ is then defined as
\begin{equation} \rho_S \,\xrightarrow{\,\,\rm{TO}\,\,}\, \Tr_{B}\, \big[ U\,\rho_S\otimes\tau\, U^\dagger \big]\,. \end{equation}
Note that thermal operations preserve the Gibbs state of the system $S$, and furthermore they obey \textit{time-translation covariance} (also  called \textit{phase-covariance} or $U(1)$-covariance), i.e. they commute with the free unitary evolution of the system: $\mathcal{T}\circ\mathcal{U}_t=\mathcal{U}_t\circ\mathcal{T}$ for any $\mathcal{T}\in\mathsf{TO}$.

\begin{figure}
\centering
\includegraphics[width=0.48\textwidth]{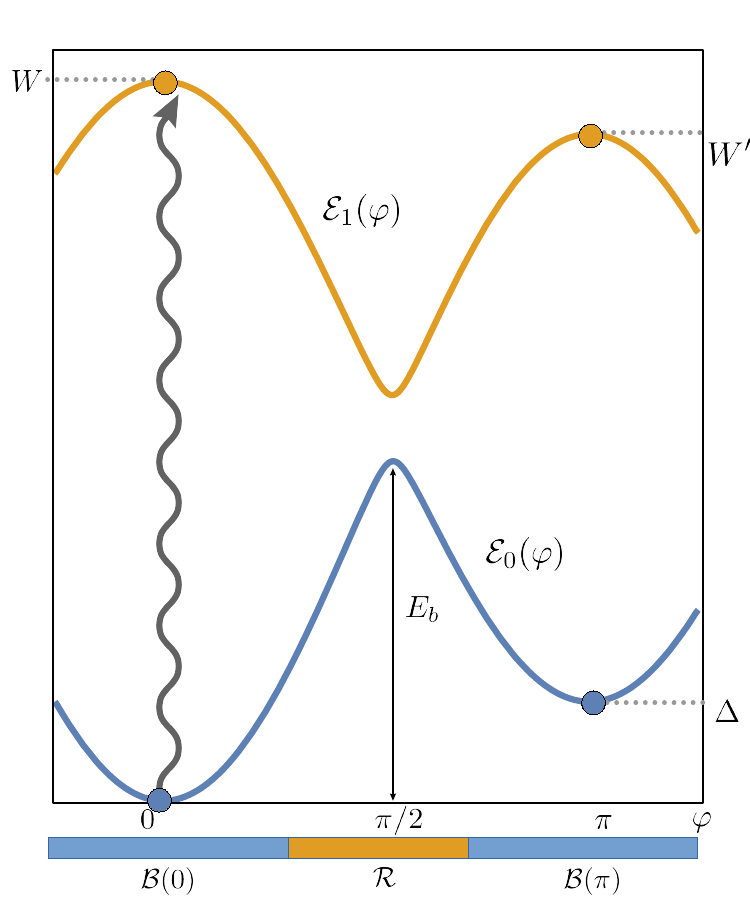}
\caption{Energy landscape for a typical photoisomer. The system starts in the electronic ground state at $\varphi=0$, and it is photoexcited (wavy arrow) by a light source. It can then relax to the cis ground state $\mathcal{E}(\varphi=\pi)$, while in contact with its environment. Our results are independent of the actual intermediate dynamics of the process. The four dots represent the states that are considered in the four-levels model of \cite{halpern2020fundamental}. As described in the main body, the two intervals $\mathcal{B}(0)$ and $\mathcal{B}(\pi)$ are introduced, defined as balls of radius $\Phi_0<\pi/2$ centered in $\varphi=0$ and $\varphi=\pi$. Then, the third interval $\mathcal{R}$ is considered, such that $\mathcal{B}(0)\cup\mathcal{R}\cup\mathcal{B}(\pi)=[0,2\pi]$.}
    \label{fig:energy_landscape}
\end{figure}

The action of thermal operations on quasiclassical states (states that are block-diagonal in the energy eigenbasis) can be fully characterised, and the associated state conversion problem, i.e. deciding whether a quasiclassical state $\rho$ can be mapped into another quasiclassical state $\sigma$ via thermal operations, can be solved via a particular version of relative majorization called \textit{thermomajorization} \cite{ruch1976principle,ruch1978mixing,horodecki2013fundamental}, in a way analogous to how standard majorization characterizes state convertibility under LOCC in the resource theory of entanglement. 
In particular, one associates to a density matrix $\rho$ a curve $L_\rho(x)$ (called \textit{thermomajorization curve}) and, given two density matrices $\rho$ and $\sigma$, it is said that $\rho$ thermomajorizes $\sigma$, i.e. $\rho\thermomaj \sigma$, if
$ L_\rho (x) \geq L_\sigma (x) \,, \forall x$.
Then, given two quasiclassical states $\rho$ and $\sigma$:
\begin{equation} 
\rho\, \xrightarrow{\mathsf{TO}}\, \sigma\,\iff\, \rho \thermomaj \sigma\,.  \end{equation}
{\emph{Photoisomerization} -- } We now consider the problem of modeling photoisomerization in the resource theory of athermality.  We will start from the same model as in \cite{halpern2020fundamental,spaventa2022capacity}, where an angular coordinate $\varphi$ between two heavy chemical groups parametrizes the relative rotation of two molecular components around a double bond. However, we will not make the simplifying assumption that the vibrational degree of freedom can only take the two stable values $\varphi=0,\pi$. Instead, we will retain the full vibrational structure and quantify its impact on the photoisomerization efficiency.  Fig.\ref{fig:energy_landscape} displays a typical energy landscape for these systems, where the two curves $\mathcal{E}_{0,1}(\varphi)$ can be obtained from a class of Hamiltonians commonly used in the study of photoisomerization (see \cite{seidner1994microscopic,seidner1995nonperturbative,hahn2000quantum,hahn2002ultrafast}). We briefly outline below how such class of Hamiltonian is typically obtained.

The simplest model that can be built for this phenomenon is the one introduced in \cite{halpern2020fundamental}, and consists in an electronic degree of freedom, modeled as a two-level system with Hamiltonian $H_{\rm el}=E_0\ket{0}\bra{0}+E_1\ket{1}\bra{1}$, coupled to a vibrational degree of freedom that serves the role of a clock. The interaction term is then chosen to be of the form typically used to describe clock-assisted dynamics \cite{Malabarba2015,woods2019autonomous,halpern2020fundamental,cilluffo2024physically}
\begin{equation}
    H_{\rm mol} = H_{\rm el}\otimes\mathbb{1}+\int_0^{2\pi} d\varphi\, V(\varphi)\otimes \ket{\varphi}\bra{\varphi}
    \label{eq:hmol}
\end{equation}
where $\ket{\varphi}$ are the angular position eigenstates for the vibrational degree of freedom. Making use of the resolution of the identity $\int_0^{2\pi} d\varphi \ket{\varphi}\bra{\varphi}=\mathbb{1}$ we can write, without loss of generality
\begin{equation}
    H_{\rm mol} = \int_0^{2\pi} d\varphi\, H_{\rm el}(\varphi)\otimes \ket{\varphi}\bra{\varphi} \,,
    \label{eq:hmol2}
\end{equation}
where $H_{\rm el}(\varphi)=H_{\rm el}+V(\varphi)$.
In order to get an expression for $H_{\rm el}(\varphi)$, one typically proceeds following \cite{seidner1994microscopic}. Given the electronic eigenstates $\ket{k}$ for $k=0,1$ introduced above, one can introduce the the \textit{torsion potentials}, $V^{(k)}(\varphi)=\bra{k}V(\varphi)\ket{k}$, which are periodic functions in $\varphi$. Then, a Fourier expansion of $V^{(k)}(\varphi)$ is taken. The result can then be exactly diagonalised, yielding the two eigenvalues $\mathcal{E}_0(\varphi)$ and $\mathcal{E}_1(\varphi)$. These two functions define the typical energy landscape for cis-trans isomerization, where the electronic ground state display a barrier while the excited state does not, allowing excited molecules to go from one configuration to the other while dissipating energy into the environment.

\begin{figure}[t]
    \centering
    \includegraphics[width=0.9\linewidth]{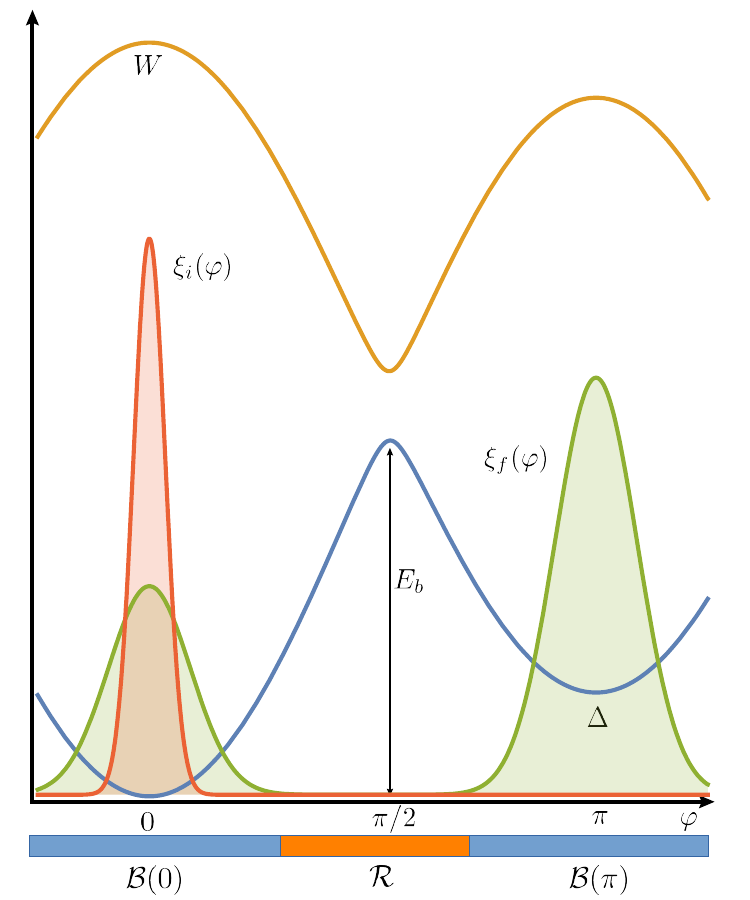}
    \caption{Visual representation of the approximations mentioned in the main body. The two intervals $\mathcal{B}(0)$ and $\mathcal{B}(\pi)$ are introduced, defined as balls of radius $\Phi_0<\pi/2$ centered in $\varphi=0$ and $\varphi=\pi$. Then, the third interval $\mathcal{R}$ is considered, such that $\mathcal{B}(0)\cup\mathcal{R}\cup\mathcal{B}(\pi)=[0,2\pi]$. The initial and final states are assumed to have angular distributions $\xi_i(\varphi)$ and $\xi_f(\varphi)$ respectively. Then, $\xi_i(\varphi)$ is assumed to only have support in $\mathcal{B}(0)$, while $\xi_f(\varphi)$ is assumed to only have support in $\mathcal{B}(0)\cup\mathcal{B}(\pi)$. The initial and final states considered in previous models are recovered in the limit $\Phi_0\to 0$.}
    \label{fig:angular}
\end{figure}

The photoisomerization yield, i.e. the efficiency of isomerization, can only be defined after a choice of which states we consider "switched". As an example, the pure state $\ket{\mathcal{E}_0(\pi)}\otimes\ket{\pi}$ can be certainly considered a \textit{trans} state, but the situation is a bit less clear for those states that have $\varphi \approx \pi/2$. This means that the definition of a photoisomerization yield implicitly defines a \textit{interval} around $\varphi=\pi$ identifying the states we considered \textit{switched}. However, if we restrict our analysis to states that are sufficiently localised in the two wells, it is natural to expect that the arbitrary choice of the extent of such an interval should not affect the results in any appreciable way. Indeed, this will turn out to be the case. To this end, let us partition the interval $[0,2\pi]$ into three parts $\mathcal{B}(0)$, $\mathcal{B}(\pi)$ and $\mathcal{R}$, as shown in Fig. \ref{fig:angular}. In particular,
$\mathcal{B}(0)$ and $\mathcal{B}(\pi)$ are intervals of radius $\Phi_0<\pi/2$ centered in $\varphi=0$ and $\varphi=\pi$ respectively, and they define the angular support of states that are considered \textit{cis} and \textit{trans}. On the other hand, $\mathcal{R}$ represents the portion of interval $[0,2\pi]$ which is not in either of the two balls. This partition induces a decomposition of the total Hilbert space into three sectors
\begin{equation}
    \mathcal{H}_{\rm mol} = \mathcal{H}_{\rm cis}\oplus \mathcal{H}_{\mathcal{R}}\oplus \mathcal{H}_{\rm trans}\,,
\end{equation}
to which we can associate the three orthogonal projectors
\begin{equation}
\begin{split}
     \Pi^{(k)}_{\rm cis} = \int_{\mathcal{B}(0)} d\varphi \,\ket{\mathcal{E}_k(\varphi)}\bra{\mathcal{E}_k(\varphi)}\otimes\ket{\varphi}\bra{\varphi}\,,\\
     \Pi^{(k)}_\mathcal{R} = \int_{\mathcal{R}} d\varphi \,\ket{\mathcal{E}_k(\varphi)}\bra{\mathcal{E}_k(\varphi)}\otimes\ket{\varphi}\bra{\varphi}\,,\\
     \Pi^{(k)}_{\rm trans} = \int_{\mathcal{B}(\pi)} d\varphi \,\ket{\mathcal{E}_k(\varphi)}\bra{\mathcal{E}_k(\varphi)}\otimes\ket{\varphi}\bra{\varphi}\,.
\end{split}
\end{equation}
Given a state $\rho$ of the molecule, we can then define the photoisomerization yield as the functional
\begin{equation}
    \gamma(\rho) = \text{Tr} \Big(\rho\, \Pi^{(0)}_{\rm trans}\Big)\,,
\end{equation}
i.e. the weight of $\rho$ on the \textit{trans} ground state. Clearly, the definition of the yield is affected by the radius $\Phi$ of $\mathcal{B}(\pi)$, i.e. by the choice of which angular configurations we consider to be \textit{trans} and which \textit{cis}. Provided that the class of states we will consider in our analysis are sufficiently localised in the two ground state minima, i.e. as long as the states are such that
\begin{equation}
    \text{Tr} \Big(\rho\, \Pi^{(0)}_{\mathcal{R}}\Big) = 0 \,,
\end{equation}
the results are independent of $\Phi$. In the technical part of this work we will provide conditions on temperature
and shape of potential that this may entail.

In order to quantify how spread out a state $\rho$ is across the angular configurations $\varphi$, we can introduce the \textit{angular distribution}
\begin{equation}
    \xi_\rho(\varphi) = \text{Tr}\Big(\rho\, \mathbb{1}\otimes \ket{\varphi}\bra{\varphi}\Big)\,.
\end{equation}
The problem of finding the maximum photoisomerization yield allowed by thermal operations can then be formulated as follows. If the dynamics is modeled by a thermal operation $\mathcal{T}\in \mathsf{TO}$ mapping a given initial state $\rho_i$ to a final state $\rho_f=\mathcal{T}(\rho_i)$, the optimal yield is defined as
\begin{equation}
    \gamma^*(\rho_i) = \sup_{\rho_i\thermomaj \rho_f} \gamma(\rho_f)\,.
\end{equation}

On the other hand, the characterization of the set of states $\rho_f$ which are thermomajorized by a given $\rho_i$, i.e. the \textit{thermal cone} of $\rho_i$, requires knowledge of the spectrum of $H_{\rm mol}$, as thermomajorization constrains the possible population transfers between energy eigenstates. However, the infinite-dimensional character of the vibrational Hilbert space renders this problem intractable, and for this reason previous models \cite{halpern2020fundamental,spaventa2022capacity} consider the vibrational degree of freedom as only taking the two values $\varphi=0,\pi$, effectively reducing it to a two level system. Then, the total Hilbert space is of dimension four, and thermomajorization can be used to compute the optimal yield allowed by thermal operations. This cutoff of the vibrational Hilbert state can be rigorously justified, under the assumption that the only initial and final states considered in the optimization are infinitely localised around the stable configurations $\varphi=0,\pi$, and have negligible weight elsewhere. This essentially corresponds to assuming that  both initial and final states have angular distributions
\begin{equation}
    \xi_{\rho_{i,f}}(\varphi) = p_{i,f} \delta(\varphi) + (1-p_{i,f})\delta(\varphi-\pi)\,,
    \label{eq:cutoff_approx}
\end{equation}
for some $p_{i,f}\in[0,1]$. Here, on the other hand, we want to characterize the entropic effect due to the availability of many vibrational configurations. Indeed, by relaxing the approximation of Eq.\eqref{eq:cutoff_approx} we are effectively allowing both initial and final states to have finite widths in $\varphi$ around the two stable configurations.
However, in order to still be able to get analytical results, we will consider states that are not too spread out in the ground state minima, in such a way to be able to work in the harmonic approximation, i.e., we will consider the functions $\mathcal{E}_{0,1}(\varphi)$ as approximately quadratic near their minima. In particular: 
\begin{equation}
\begin{split}
&\mathcal{E}_0(\varphi) \approx \frac{1}{2} I \omega_0^2 \varphi^2 \text{ when } \varphi\approx 0 \,, \\   
&\mathcal{E}_0(\varphi) \approx \Delta+\frac{1}{2} I \omega_\Delta^2 (\varphi-\pi)^2 \text{ when } \varphi\approx \pi\,,\\
&\mathcal{E}_1(\varphi) \approx \mathcal{E}_1(\pi/2) +\frac{1}{2} I \omega_b^2 (\varphi-\pi/2)^2 \text{ when } \varphi\approx \pi/2\,,
\end{split}
\label{eq:energy_expansion}
\end{equation}
where we have defined
\begin{equation}
    I \omega_{0,\Delta}^2 \equiv \frac{d^2\mathcal{E}_0(\varphi)}{d\varphi^2}\Big\rvert_{\varphi=0,\pi}\quad \text{and}\quad  I\omega_{b}^2 \equiv \frac{d^2\mathcal{E}_1(\varphi)}{d\varphi^2}\Big\rvert_{\varphi=\pi/2}\,,
\end{equation}
and $I$ is the moment of inertia associated to the angular degree of freedom $\varphi$. Without loss of generality, we can reabsorb this quantity in the definitions of $\omega_{0,\Delta,b}$, or equivalently, set $I\equiv 1$.\\

{\emph{Static solution} -- } In this section we compute the optimal photoisomerization yield under the assumption that the molecule can be described via the Hamiltonian in Eq.\eqref{eq:hmol}, which can be interpreted as a hybrid Hamiltonian \cite{diosi2023hybrid} for the hybrid quantum-classical system of a quantum two-level system coupled to a classical angular degree of freedom. In other words, this equates to considering the angular degree of freedom $\varphi$ as a classical label whose role is merely that of selecting the correct electronic Hamiltonian $H_{\rm el}(\varphi)$. This is equivalently expressed by the fact that the eigenstates of this Hamiltonian are trivially of the form
\begin{equation}
   \ket{\psi_j(\varphi)} = \ket{\mathcal{E}_j(\varphi)}\otimes \ket{\varphi}\,,
\end{equation}
corresponding to the eigenenergies $\mathcal{E}_j(\varphi)$, since
\begin{eqnarray}
     H_{\rm mol} \ket{\psi_j(\varphi)} &=& \int_0^{2\pi} d\varphi'\, \bra{\varphi'}\ket{\varphi} H_{\rm el}(\varphi')\ket{\mathcal{E}_j(\varphi)} \otimes \ket{\varphi'}\nonumber\\
     &=& (H_{\rm el}(\varphi)\otimes \mathbb{1} )\ket{\mathcal{E}_j(\varphi)} \otimes \ket{\varphi}\nonumber\\ 
     &=& \mathcal{E}_j(\varphi) \ket{\psi_j(\varphi)}\,.
\end{eqnarray}
We will focus on quasiclassical states parametrized as
\begin{equation}
    \rho = \sum_{k}\int_0^{2\pi}d\varphi\, p_{k}(\varphi) \ket{\mathcal{E}_k(\varphi)}\bra{\mathcal{E}_k(\varphi)}\otimes \ket{\varphi}\bra{\varphi}\,,
\end{equation}
where the probability distribution $p_k(\varphi)$ uniquely identify the weight on each eigenstate $\ket{\psi_j(\varphi)}$. Then, the photoisomerization yield takes the form
\begin{equation}
    \gamma = \int_{\mathcal{B}(\pi)} d\varphi\, p_0(\varphi)\,.
    \label{eq:yield_def}
\end{equation}

In \cite{halpern2020fundamental,spaventa2022capacity} both the initial and final states are assumed to only have finite weight on the vibrational configurations $\varphi=0,\pi$. On the other hand, we introduce here the possibility of having distributions $p_k(\varphi)$ on different angular configurations. 
Let us consider an initial state $\rho_i$ associated with

\begin{equation}
    p_k(\varphi) = \Big( q\delta_{k,1} + (1-q)\delta_{k,0}\Big)\frac{1}{\sqrt{2\pi}\sigma}e^{-\frac{\varphi^2}{2\sigma^2}}\,,
\end{equation}
that is, a Gaussian population distribution over angular configurations, of which a fraction $q$ has been excited to the electronic state $k=1$, and a fraction $1-q$ remains in the electronic ground state. This is a generalization of the previous models that takes into account a finite width of the distribution over angular configurations. In particular, the initial state considered in the previous models of \cite{halpern2020fundamental,spaventa2022capacity} is recovered in the limit $\sigma\to 0$.
Given the expansion of Eq.\eqref{eq:energy_expansion} as a quadratic function of $\varphi$, and as in our static picture the 
quantum uncertainties are negligible, the expression above defines the initial state as a statistical mixture of two Gaussian distributions (one in the ground state manifold and one in the excited state manifold) with effective inverse temperature 
\begin{equation}
    \tilde\beta=\frac{1}{\sigma^2 \omega_0^2}\,.
    \label{eq_betaeff}
\end{equation}
We can then distinguish two cases, depending on whether $\tilde\beta>\beta$ or $\tilde\beta\leq\beta$. These two conditions correspond to two classes of initial states, whose effective temperatures are larger or smaller than the background temperature $\beta$. In keeping consistency with previous approaches to the problem, we will here focus on the case $\tilde{\beta}>\beta$, which is the scenario that includes the limit $\tilde{\beta}\to \infty$ in which infinitely localised states are recovered. The other scenario involves states whose support is rather spread out, and are therefore in contrast with the assumptions and approximations made so far. \\

In order to make use of the thermomajorization partial order, around each of the two ground state minima the vibrational degree of freedom $\varphi$ is discretized on a lattice composed of $N$ points $\varphi_k = k\varphi_0$ with lattice constant $\varphi_0 = 2\pi/N$. We will impose the thermomajorization condition on states in this discretized Hilbert space and then perform the continuum limit $\varphi_0 \to 0$.
In this way, by exploiting the Gaussianity of the initial state, we can single out the $\beta$-ordering of the initial state, which determines the thermomajorization curve for the state of the photoexcited system. The explicit construction of this curve is detailed in Appendix \ref{appendix_continuum}. 
In order to compute $\gamma^*$, we construct the thermomajorization curve of the final state by imposing two constraints: a) such a curve must lie below the curve associated to the initial state, and b) the yield associated with such curve must be as high as possible.
The optimal yield in this case is found to be
\begin{eqnarray}
    \gamma^*_{\rm stat} &=& q\\
    && \hspace*{-0.75cm}+(1-q)\,\text{erf}\left[\sqrt{\frac{\tilde\beta}{\beta}}\text{erf}^{-1}\left(\frac{\omega_0}{\omega_\Delta}e^{-\beta\Delta} - \sqrt{\frac{\omega_0^2\beta}{2\pi}}e^{-\beta W}\right)\right]\,.\nonumber
\end{eqnarray}

\begin{figure}[t]
    \centering
    \includegraphics[width=0.9\linewidth]{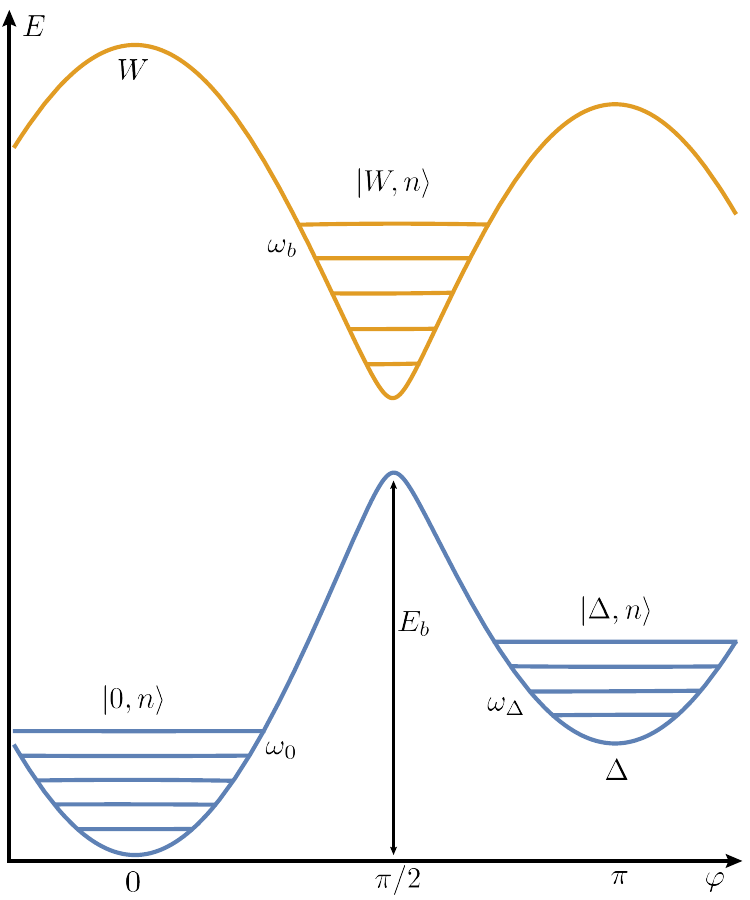}
    \caption{Schematics for the symbols used in the main body. Near each energy landscape minima, the curves $\mathcal{E}_k(\varphi)$ are approximately quadratic. Then, each of these harmonic wells (corresponding to $\varphi=0,\pi/2,\pi$) gives rise to a sequence of well-localised energy levels, with gaps denoted by $\omega_0, \omega_b, \omega_\Delta$ respectively. The harmonic oscillator eigenstates corresponding to the cis ground state are denoted by $\ket{0,n}$ for $n=0,1,2,\dots$, and the same for the other wells. In the limit of the oscillator frequencies being much smaller than the barrier energy $E_b$, the number of levels in each well is very large, and the thermomajorization curves associated to states of the system can be approximated by smooth concave functions.}
    \label{fig:spectrum}
\end{figure}

{\emph{Impact of vibrational dynamics} -- } Here we extend the results so far by explicitly adding a kinetic contribution, associated to the vibrational degree of freedom $\varphi$, to the total Hamiltonian $H_{\rm mol}$. The new term has the form
\begin{equation}
    T_\varphi = -\frac{\hbar^2}{2 I} \frac{\partial^2}{\partial \varphi^2}\,, 
\end{equation}
where we can once again set $I\equiv 1$, and leads to the fact that the energy eigenstates of the composite system are not of the form $\ket{\mathcal{E}(\varphi)}\otimes\ket{\varphi}$ anymore, and are in general difficult to compute. However, we can perform the following approximation. Let us again consider initial and final states that are sufficiently localised in the two stable configurations $\varphi=0$ and $\varphi=\pi$. This means that the only levels which have a non-vanishing population are those with wavefunctions that have negligible overlap with angular configurations far from the two stable ones, where the potential associated with the electronic ground state, $\mathcal{E}_0(\varphi)$, is approximately quadratic. Thus, the only eigenstates of $H_{\rm mol}$ that we are interested in are those arising from two harmonic potentials centered at $\varphi=0,\pi$, i.e. equally spaced levels associated to states well localised in each harmonic well. This will be the case as long as the energy spacing between harmonic levels is small compared to the barrier energy, i.e.
\begin{equation}
    \frac{\hbar \omega_0}{E_b} \ll 1\quad \mbox{and} \quad \frac{\hbar \omega_\Delta}{E_b-\Delta} \ll 1\,,
    \label{eq:conditions}
\end{equation}
and as long as the effective temperature of Eq. \eqref{eq_betaeff} is such that 
\begin{equation}
    \tilde{\beta} E_b \ll 1\,.
\end{equation}
The first two conditions are certainly satisfied in practice, while the second one restrict the family of states that we consider in our analysis to those whose population distributions over eigenstates does not involve levels with energy comparable to $E_b$. Since we are interested in the optimal photoisomerization yield, this assumption will not impact the results much, because populating levels with high energy is a rather inefficient way to increase the yield. The validity of this approximation can be further strengthened by considering the concept of \textit{transness} as defined in \cite{chuang2022steady}. There, the authors use it to quantify the localization of the photoisomer's eigenstates, and what they find is that eigenstates with energy lower than the barrier are very well localised in the two minima. In the excited state manifold, we can perform a harmonic approximation near $\varphi=\pi/2$, and consider the initial distribution as a superposition of such levels. However, because $\beta E_b, \beta W\gg 1$, the internal distribution of population over such levels does not affect the thermomajorization curve in a relevant way. Indeed, the only relevant figure of merit is the sum of all populations in these levels, which we parametrize again as $q\in[0,1]$. The spacing between levels is determined by the harmonic potential's width in the expansion of Eq. \eqref{eq:energy_expansion}, as in the previous case. The situation is depicted in Fig. \ref{fig:spectrum}.

In order to simplify the notation, let us denote the energy eigenstates by $\ket{E,n}$, where $E=0,\Delta,W$ specifies the energy landscape minimum and $n$ denote the phonon number in that harmonic potential. Due to the high degree of localization of low energy eigenstates in the \textit{trans} well, the states $\ket{\Delta,n}$ can all be expressed as linear combinations of states $\ket{\mathcal{E}_0(\varphi)}\otimes\ket{\varphi}$ corresponding to $\varphi\in\mathcal{B}(\pi)$ only. This, in turn, means that the projector $\Pi_{\rm trans}^{(0)}$ used to define the yield can be rewritten via a local change of basis for $\mathcal{H}_{\rm trans}^{(0)}$, so that the yield now reads
\begin{equation}
    \gamma(\rho) = \text{Tr} \sum_n\Big( \rho \ket{\Delta,n}\bra{\Delta,n} \Big) =\sum_n \bra{\Delta,n}\rho\ket{\Delta,n}\,,
\end{equation}
i.e. the sum of weights on the energy eigenstates localised in the \textit{trans} ground state well, as one expects.\\

In keeping the same spirit as the previous scenario, let us consider, for $0\leq q \leq 1$, an initial state of the form
\begin{equation}
    \rho_i = q \ket{W,0}\bra{W,0}+\frac{1-q}{Z_0(\tilde{\beta})}\sum_n e^{-\tilde{\beta}\hbar\omega_0 n}\ket{0,n}\bra{0,n}\,,
\end{equation}
where $Z_0(\tilde{\beta})=\sum_n e^{-\tilde{\beta}\hbar\omega_0 n} = \frac{1}{1-e^{-\tilde{\beta}\hbar\omega_0 }}$ and $\tilde{\beta}$ is an effective temperature that parametrizes the population distribution. We will consider the case $\tilde{\beta}>\beta$, i.e. when the initial state is effectively colder than its thermal equivalent at background temperature $\beta$, and furthermore, since $W$ is generally very large, we again consider the case $q\geq\tilde{q}$ so that the $\beta$-ordering of the initial state is fixed as $\{\ket{W,0},\ket{0,0},\ket{0,1},\ket{0,2},\dots\}$. However, in order to find a closed form expression for the optimal yield, we make use of the conditions introduced earlier in Eq.\eqref{eq:conditions}. In particular, under this approximation, the thermomajorization curve of the initial state can be approximated with a smooth concave function defined as (see Appendix \ref{appendix_dynamic} for details)
\begin{equation}
    \mathcal{L}_i(x) =  q + (1-q) \Big( 1 - \big( 1 - (1-e^{-\beta \hbar \omega_0})x \big) ^{\tilde{\beta}/\beta} \Big) \,.
\end{equation}
The corresponding optimal yield can be found by setting $x+e^{-\beta W}=Z_\Delta(\beta)$
\begin{equation}
    \gamma^*_{\rm dyn} = \mathcal{L}_i\left(\frac{e^{-\beta \Delta}}{1-e^{-\beta\hbar\omega_\Delta}}-e^{-\beta W}\right)
\end{equation}
which leads to
\begin{eqnarray}
    \gamma^*_{\rm dyn} &=& q\\
    && \hspace*{-1.cm}+(1-q) \left[1{-}\left[1{-}\left(1{-}e^{-\beta\omega_0}\right) \left(\frac{e^{-\beta  \Delta }}{1{-}e^{-\beta \omega_\Delta}}{-}e^{-\beta W }\right)\right]^{\frac{\tilde{\beta}}{\beta} }\right]\,.\nonumber
\end{eqnarray}

\begin{figure}
    \centering
    \includegraphics[width=0.95\linewidth]{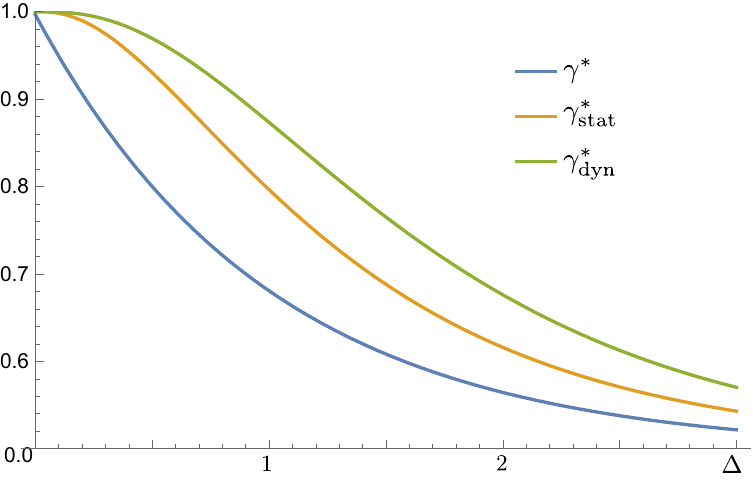}
    \caption{Comparison of the two solutions $\gamma^*_{\rm stat}$ and $\gamma^*_{\rm dyn}$ with the optimal yield $\gamma^*$ obtained in the 3-level model. Here $W=5$, $q=0.5$, $\omega_0=\omega_\Delta=0.1$, $\beta=1$, and $\tilde{\beta}=3$.}
    \label{fig:comparison}
\vspace{2ex}
    \centering
    \includegraphics[width=0.95\linewidth]{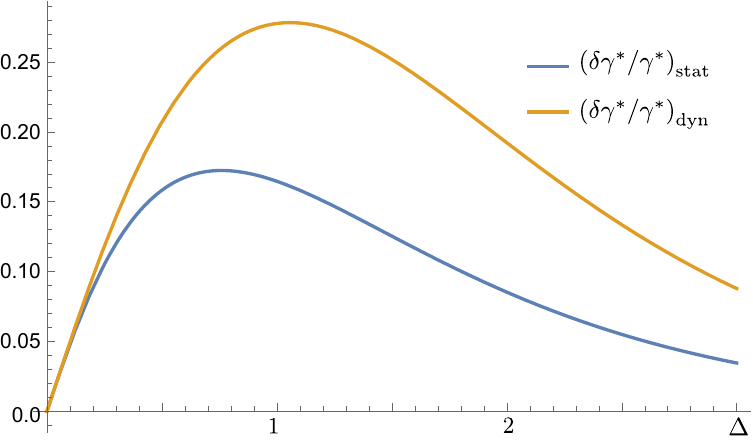}
    \caption{Relative advantage for the photoisomerization yield with respect to the optimal solution $\gamma^*$. Here $W=5$, $q=0.5$, $\omega_0=\omega_\Delta=0.1$, $\beta=1$, and $\tilde{\beta}=3$.}
    \label{fig:advantage}
\end{figure}

{\emph{Discussion} -- } The solution obtained above and in the previous sections are compared in Fig.\ref{fig:comparison}. We can see that in both scenarios (i.e. for both static and dynamic solutions) the availability of many angular configurations has an entropic effect that boosts the photoimerization yield. However, when the kinetic contribution to the vibrational Hamiltonian is present, the yield increase is larger. It is not immediately clear why the presence of the kinetic term should be beneficial for the yield, but the following simple example offers an intuitive explanation of the effect. Consider a harmonic oscillator with Hamiltonian $H=\frac{1}{2}m\omega x^2 + \frac{p^2}{2 m}$. The partition function for this system is $Z= \frac{1}{2}\text{cosech}(\beta\omega/2)$, and its corresponding equilibrium free energy $F = -\frac{1}{\beta}\ln\left( \frac{1}{2}\text{cosech}(\beta\omega/2) \right) $. If one neglects the kinetic term, i.e. $H \approx H_0 = \frac{1}{2}m\omega x^2$, one has $Z_0(\beta) = \sqrt{\frac{2\pi}{\beta m\omega^2}}$ and $F_0 = -\frac{1}{2\beta}\ln\left(\frac{2\pi}{m\beta \omega^2}\right) $. For $\beta\omega$ large enough, one has $F>F_0$, and we conclude that the presence of the kinetic term effectively translates into a larger free energy. In order to quantify the deviations from the 3-level model solution
\begin{equation}
    \gamma^* = q + (1-q)(e^{-\beta \Delta}-e^{-\beta W})\,,
\end{equation}
we can introduce the relative advantages
\begin{equation}
    \left(\frac{\delta\gamma^*}{\gamma^*}\right)_{\rm stat} = \frac{\gamma^*_{\rm stat}-\gamma^*}{\gamma^*}\,,\quad \left(\frac{\delta\gamma^*}{\gamma^*}\right)_{\rm dyn} = \frac{\gamma^*_{\rm dyn}-\gamma^*}{\gamma^*}\,.
\end{equation}
The relative advantages are plotted in Fig.\ref{fig:advantage}.

We can conclude that the thermomajorization techniques used to bound physical quantities under thermal operations, can still be deployed for systems that go beyond qubits and qutrits, and instead have large (or possibly infinite) dimensionalities, provided that one introduces the necessary approximations. It would be then interesting to repeat the analysis of \cite{spaventa2022capacity} to find the optimal yield under the additional assumption of Markovianity, so as to bound the impact of memory effects in this less artificial model. However, Markovian thermomajorization \cite{lostaglio2017markovian} is a procedure that doesn't have an efficient scaling in the Hilbert space dimension $d$. In particular, deciding whether two states continuously thermomajorize each other require the construction of $O(d!)$ intermediate thermomajorization curves, which is impractical for larger systems ($d>6$), even numerically. Another interesting direction would be considering two molecules prepared in an entangled state and quantify the impact of quantum coherence as in \cite{burkhard2023boosting}. Finally, the optimal yield for multiple ($N>2$) photoswitches would be surely worth investigating, to see how the efficiency would approach the one given by the standard (asymptotic) thermodynamic regime, and also to find bounds on the impact of intermolecular correlations on the efficiency of photoisomerization.

{\em Acknowledgements:} This work was supported by the ERC Synergy grant HyperQ (grant no. 856432) and the QuantERA project ExTRaQT (grant no. 499241080).

\bibliographystyle{unsrt}
\bibliography{ms.bib}


\onecolumngrid
\appendix
\section{Static solution}
\label{appendix_continuum}
\subsection{Setup}

In this section we provide details on the calculation of the optimal yield in the static scenario described by the Hamiltonian
\begin{equation}
    H_{\rm mol} = H_{\rm el}\otimes\mathbb{1}+\int_0^{2\pi} d\varphi\, V(\varphi) \otimes \ket{\varphi}\bra{\varphi}\,.
\end{equation}
We expand the electronic ground state eigenvalue near the two stable configurations as specified by Eqs. \eqref{eq:energy_expansion}, and then we discretize the interval $[0,2\pi]$ in $N$ points $\{ n\varphi_0 \}_{n=0}^N$ with lattice spacing $\varphi_0=2\pi/N$. The harmonic expansion requires another assumption, for consistency reasons: the initial and final states we are going to consider in our analysis have negligible support on vibrational subspaces that are very far from the stable configurations. In fact, if this was not the case, the state support would include intervals in $\varphi$ for which the deviation form harmonicity are substantial. The harmonic approximation can be relaxed without affecting the crux of the following technique, but doing so would preclude the simplicity of the closed-form analytical solution. 

\subsection{Constructing the Curve for the Initial State}
We assume that initially the system has negligible weight outside of $\mathcal{B}(0)$. In particular, as described in the main body, let us consider an initial state whose weight on the discretized states $\ket{\mathcal{E}_k(\varphi_n)}\otimes\ket{\varphi_n}$ for $\varphi_n \in \mathcal{B}(0)$ is specified by the distribution 
\begin{equation}
    p_k(\varphi_n) = \Big( q\delta_{k,1} + (1-q)\delta_{k,0}\Big)f(\varphi_n)\,,
\end{equation}
i.e. a distribution $f(\varphi_n)$ in the ground state manifold of which a fraction $q$ has been excited to the excited state manifold. For convenience of notation, we introduce the integer $M=\max\{n \,|\, \varphi_n \in \mathcal{B}(0)\}$ as the index defining the boundary of $\mathcal{B}(0)$. We then take a Gaussian profile for $f$, which reads
\begin{equation}
f_n \equiv f(\varphi_n) = \frac{1}{Z_0(\tilde{\beta})}e^{-\tilde{\beta}\omega^2_0\varphi_0^2 n^2/2}\,, \quad \text{where}\quad Z_0(\tilde{\beta})=2\sum_{n=0}^{M} e^{-\tilde{\beta}\omega^2_0\varphi_0^2 n^2/2}\,,
\end{equation}
Thanks to the harmonic approximation near the minima, we can interpret Gaussian distributions as being thermal distributions, but at an effective inverse temperature $\tilde{\beta}$ that can be distinct from the environmental $\beta$.

In the continuum limit $\varphi_0 \ll 1$, we can define the continuous variable $\varphi\equiv n\varphi_0$ and the maximum angle $\Phi\equiv M\varphi_0$, and we get 
\begin{equation}
    f_n \longmapsto f(\varphi) = \frac{1}{Z_0(\tilde{\beta})}e^{-\tilde{\beta}\omega^2_0\varphi^2/2}\,, \quad \text{where}\quad Z_0(\tilde{\beta})\longmapsto \frac{2}{\varphi_0}\int_0^{\Phi} e^{-\tilde{\beta}\omega^2_0\varphi^2/2} d\varphi = \sqrt{\frac{2 \pi}{\tilde{\beta}\omega^2_0 \varphi_0^2}}\text{erf}\left(\sqrt{\frac{\tilde{\beta}\omega^2_0}{2}}\Phi\right).
\end{equation}
Furthermore, if $\tilde{\beta}$ is large enough, meaning that the initial state's effective temperature is low enough, we can safely extend the integral from $0$ to $\Phi\to\infty$ and therefore
\begin{equation}
    \quad Z_0(\tilde{\beta})\longmapsto \sqrt{\frac{2 \pi}{\tilde{\beta}\omega^2_0 \varphi_0^2}}\,.
\end{equation}

To construct the thermomajorization curve associated to the initial state, we need to find its $\beta$-ordering, i.e. we need to consider its population vector in the energy eigenbasis and arrange in decreasing order the quantities $p_ie^{\beta E_i}$, formed by ratios of populations of the $i$-th energy level and their corresponding Boltzmann factors $e^{-\beta E_i}$. The first possible simplification is the following: given that typically $W\ll \Delta$, states with corresponding to the excited state manifold will always come first in the $\beta$-ordering, as long as $q$ is finite. Furthermore, the internal ordering of those states in the excited manifold doesn't matter, because when put together to construct the thermomajorization curve, they occupy an overall interval of order $o(e^{-\beta W})$ on the horizontal axis. Therefore, without loss of generality, we can consider the initial state distribution as being
\begin{equation}
    p_k(\varphi_n) =  q\delta_{k,1}\delta_{n,0}  + (1-q)\delta_{k,0}f_n\,,
\end{equation}
i.e. by concentrating all the populations over the excited state manifold into the single state $\ket{\mathcal{E}_1(\varphi_0)}\otimes\ket{\varphi_0}$ corresponding to energy $W$. Therefore, the first elbow point of the thermomajorization curve associated to the initial state is the point $(e^{-\beta W},q)$ on the $y$-axis.

As per the other populations, the assumption of Gaussianity allows us to find their $\beta$-order in terms of a single parameter. Indeed, their $\beta$-ordering is uniquely specified by the sign of $\beta-\tilde{\beta}$. In particular, consider the case where the initial cis ground state distribution is effectively cooler than the environment, i.e., $\tilde{\beta}\geq\beta$.
The arrangement of the terms $(1-q) e^{(\beta-\tilde{\beta})k_0n^2\varphi_0^2}$ in decreasing order,
in the limit $\varphi_0\to 0$, can be expressed via a compact, integral representation. In particular, the $x$-coordinates of elbow points corresponding to the first $M+1$ levels, may be rewritten as
\begin{equation}
    x_n \approx e^{-\beta W}+\int_{-n\varphi_0}^{n\varphi_0}  \frac{d\varphi}{\varphi_0}\, e^{-\beta \omega^2_0 \varphi^2/2} \,, \quad\text{if}\quad n\leq M\,,
\end{equation}
while the $x$-coordinates for the last $M$ levels read
\begin{equation}
    x_n = e^{-\beta W} + x_{M} + \int_{-n\varphi_0}^{n\varphi_0}  \frac{d\varphi}{\varphi_0}\, e^{-\beta (\Delta+ \omega^2_\Delta \varphi^2/2)}\,, \quad\text{if}\quad n> M\,.
\end{equation}
Meanwhile, the corresponding $y$-coordinates are

\begin{equation}
    y_n = q + \frac{1-q}{Z_0(\tilde{\beta})}\int_{-n\varphi_0}^{n\varphi_0}  \frac{d\varphi}{\varphi_0}\, e^{-\tilde{\beta} \omega^2_0 \varphi^2/2} \,, \quad\text{if}\quad n\leq M\,,
\end{equation}
and
\begin{equation}
    y_n = 1\,,\quad\text{if}\quad n> M\,.
\end{equation}

\begin{figure}[h]
    \centering
    \includegraphics[width=0.9\linewidth]{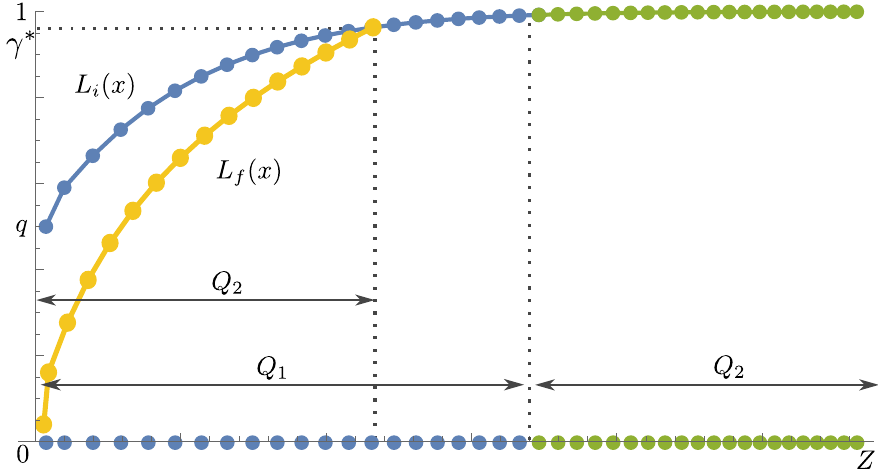}
    \caption{Illustrative sketch for the construction of the thermomajorization curves and the computation of the optimal yield. The curve $L_i(x)$ corresponding to the initial state displays the partition in blue and green segments, as described in the main body. The total lengths $Q_1$ and $Q_2$ are indicated on the horizontal axis. A possible final thermomajorization curve $L_f(x)$ is shown in orange. It touches the initial curve in one point and then it can have any functional behaviour, provided that it stays under $L_i(x)$. The optimal yield $\gamma^*$ is then computed as the value of the initial state curve on the horizontal coordinate $x=Q_2$.}
    \label{fig:thermo_curves}
\end{figure}

    
Fig.(\ref{fig:thermo_curves}) shows a sketch of the resultant initial thermomajorisation curve. Call the curve $y=L_i(x)$.

\subsection{Optimising the Yield: The Curve for the Final State}
In this section we construct the thermomajorization curve associated to the final state: i.e., any valid thermomajorisation curve that sits below that of the initial state, and achieves the maximum possible weight over the states around $\Delta$, thus maximising the yield. When constructing the curve, which is a piecewise linear curve formed by joining segments associated to energy eigenstates, we can consider the two groups of segments associated to each harmonic well. 

For convenience, call (and colour) the segments corresponding to energy levels situated around $\ket{0}$, `blue'. In other words, blue segments are those of lengths $e^{-\beta \omega_0^2 n^2\varphi_0^2/2}$. Call and colour the segments corresponding to the energy levels associated with the minimum at $\ket{\Delta}$, `green'. Green segments have lengths $e^{-\beta(\Delta+\omega_\Delta^2 n^2\varphi_0^2/2)}$. A relevant figure of merit is then the total horizontal increment associated to the blue/green segments, defined as the measure of the support of the blue/green segments alone. The total length of the blue portion, in the continuum limit $\varphi_0\ll 1$, can be represented as the integral
\begin{equation}
    Q_1 = \int_{-\Phi}^{\Phi}  \frac{d\varphi}{\varphi_0}\,e^{-\beta \omega^2_0 \varphi^2/2}\,,
\end{equation}
 while the green portion stretches across a total of
 \begin{equation}
    Q_2=\int_{-\Phi}^{\Phi} \frac{d\varphi}{\varphi_0}\, e^{-\beta(\Delta+ \omega^2_\Delta \varphi^2/2)}\,.
\end{equation}
The optimal yield is then simply the total vertical increment of the thermomajorisation curve over the length spanned by green segments. A curve arising from a rearrangement of blue and green segments is a valid thermomajorisation curve as long as it is concave. Our goal is then to construct a final thermomajorization curve by (i) rearranging the blue and green segments, and (ii) maximizing the rise of the curve over the green segments.
Given the concavity of the curve, its slope is larger for smaller values of $x$. In order to maximise the vertical increment over green segments, these should thus be clustered together, right underneath the region where the slope of the curve is largest, i.e., to the left. This is precisely what is shown in Fig.(\ref{fig:thermo_curves}). The blue segments follow, and their ordering is not relevant for our calculation. Now, clearly the maximum possible height gain of the final curve over the green segments is the value of the initial curve, otherwise the initial state would not thermomajorize the final one. Therefore, the optimal yield reads
 The optimal yield is then computed as
\begin{equation}
    \gamma^* = L_i (x)\Bigr|_{x=Q_2} = f_i \left( \int_{-\Phi}^{\Phi} \frac{d\varphi}{\varphi_0} e^{-\beta(\Delta+ \omega^2_\Delta \varphi^2/2)} \right)\,,
\end{equation}
i.e., the $y$-coordinate of the initial thermomajorization curve at $x=Q_2$, the total length of the green stretch.\\

To find the value of the curve $L_i$ at the coordinate $x=Q_2$, we need to compute the number $p$ of blue segments required to span the length $Q_2$. In other words, we need to find the integer $p$ that solves the equation
\begin{equation}
    x_p = Q_2\,,
\end{equation}
which can be rewritten as
\begin{equation} 
    e^{-\beta\Delta}\int_{-\Phi}^{\Phi} d\varphi\, e^{-\beta \omega^2_\Delta  \varphi^2/2} 
    = e^{-\beta W}+\int_{-p\varphi_0}^{p\varphi_0}d\varphi\, e^{-\omega^2_0 \beta \varphi^2/2}\,.
\label{eq:p_constr}
\end{equation}
Given the solution $p$ to the equation above, we can finally write the optimal yield as
\begin{equation}
    \gamma^* = q + (1-q)\frac{\int_{-p\varphi_0}^{p\varphi_0} d\varphi\, e^{-\omega^2_0\tilde\beta\varphi^2/2}}{\int_{-\Phi}^{\Phi}d\varphi\, e^{- \omega^2_0\tilde\beta\varphi^2/2}} = q + \frac{1-q}{Z_0(\tilde{\beta})}\sqrt{\frac{2\pi}{\tilde{\beta}\omega^2_0}}\text{erf}\left( \sqrt{\frac{\tilde{\beta} \omega^2_0}{2}} \frac{\varphi_0}{2}p \right)\,.
\end{equation}
Under the assumption that the Gaussian distributions are well-localised, we can extend the integration from $[-\Phi,\Phi]$ to $[-\infty,+\infty]$. Then, Eq.\eqref{eq:p_constr} reduces to 
\begin{equation}
    \int_{-p\varphi_0}^{p \varphi_0}d\varphi\, e^{-\omega^2_0 \beta \varphi^2/2} 
    = e^{-\beta\Delta}\sqrt{\frac{2\pi}{\beta\omega^2_0}}-e^{-\beta W}\,,
\end{equation}
which has solution
\begin{equation}
    p = \frac{1}{\varphi_0}\sqrt{\frac{2}{\beta \omega_0^2}}\text{erf}^{-1}\left( \frac{\omega^2_0}{\omega^2_\Delta}e^{-\beta \Delta}-\sqrt{\frac{\beta\omega^2_0}{2\pi}}e^{-\beta W} \right)\,,
\end{equation}
and the optimal yield reads
\begin{equation}
\gamma^*=q+(1-q)\,\text{erf}\left[\sqrt{\frac{\tilde\beta}{\beta}}\text{erf}^{-1}\left( \frac{\omega^2_0}{\omega^2_\Delta} e^{-\beta\Delta} - \sqrt{\frac{ \beta\omega^2_0}{2\pi}}e^{-\beta W}\right)\right]\,.
\end{equation}

\section{Entropic effect of the vibrational dynamics}
\label{appendix_dynamic}

In this section we provide details on the calculation of the optimal yield in the scenario described by the Hamiltonian
\begin{equation}
    H_{\rm mol} = H_{\rm el}\otimes\mathbb{1}+\int_0^{2\pi} d\varphi\, V(\varphi) \otimes \ket{\varphi}\bra{\varphi} -\mathbb{1}\otimes\frac{\hbar^2}{2 I} \frac{\partial^2}{\partial \varphi^2}\,,
\end{equation}
i.e., when a kinetic term for the vibrational d.o.f. $\varphi$ is added. Following the arguments presented in the main body, the eigenstates of $H_{\rm mol}$ can be approximated as those arising from the harmonic minima, and are denoted by $\ket{E,n}$, for $E=0,\Delta,W$. The widths of the harmonic wells define the frequencies $\omega_0,\omega_\Delta,\omega_b$ as in Eq.\eqref{eq:energy_expansion}. The initial state is taken as 
\begin{equation}
    \rho_i = q \ket{W,0}\bra{W,0}+\frac{1-q}{Z_0(\tilde{\beta})}\sum_n e^{-\tilde{\beta}\hbar\omega_0 n}\ket{0,n}\bra{0,n}\,,
\end{equation}
where $Z_0(\tilde{\beta})=\sum_n e^{-\tilde{\beta}\hbar\omega_0 n} = \frac{1}{1-e^{-\tilde{\beta}\hbar\omega_0 }}$ and $\tilde{\beta}>\beta$ is an effective temperature that parametrizes the population distribution. Once again, due to the fact that $W$ is generally very large, we can focus w.l.o.g. on the case $q\geq\tilde{q}=\frac{1}{1+e^{\beta W}}\approx 0$ so that the $\beta$-ordering of the initial state is fixed as $\{\ket{W,0},\ket{0,0},\ket{0,1},\ket{0,2},\dots\}$. \\

Now, the thermomajorization curve $L_i(x)$ is then defined as the linear interpolation of the points
\begin{equation}
\begin{split}
    &P_{-2} = (0,0)\,,\quad P_{-1}=(e^{-\beta W},q)\,,\\
    &P_{k}=(x_k,y_k)\,,\quad k=0,1,2,\dots
\end{split}
\end{equation}
where
\begin{equation}
\begin{split}
    & x_k =  e^{-\beta W}+\sum_{n=0}^k e^{-\beta \hbar \omega_0 n} = e^{-\beta W}+\frac{1-e^{-\beta\hbar\omega_0 (k+1)}}{1-e^{-\beta\hbar\omega_0}}\,,\\
    & y_k = q+\frac{1-q}{Z_0(\tilde{\beta})}\sum_{n=0}^k e^{-\tilde{\beta} \hbar \omega_0 n} = q +(1-q)\left(1-e^{-\tilde{\beta}\hbar\omega_0 (k+1)}\right)\,.
\end{split}
\end{equation}
Thus, the thermomajorization curve of the initial state is constructed by joining the interpolating segments 
\begin{equation}
    s_k(x) = y_{k-1} + \frac{y_k-y_{k-1}}{x_k-x_{k-1}}(x-x_{k-1})
\end{equation}
with domain $\mathcal{D}_k=[x_{k-1},x_k]$, connecting points $P_{k-1}$ and $P_{k}$. We then define a mapping 
$$k^*:\bigcup_k \mathcal{D}_k \to \mathbb{N}$$ 
that maps each $x\in \bigcup_k \mathcal{D}_k$ to the value $k^*$ such that $x\in \mathcal{D}_{k*}$. A possible expression for such a function is
\begin{equation}
    k^*(x) = \sum_k \theta(x-x_k)\,.
\end{equation}
Finally, the thermomajorization curve associated with the initial state reads
\begin{equation}
    L_i(x)= s_{k^*(x)}(x)\,,
\end{equation}
and the optimal yield is then computed as
\begin{equation}
    \gamma^* = L_i(x=Z_\Delta(\beta))\,,
\end{equation}
where
\begin{equation}
    Z_\Delta(\beta)=\sum_{n=0}^\infty
e^{-\beta (\Delta + \hbar\omega_\Delta n)} = \frac{e^{-\beta \Delta}}{1-e^{-\beta\hbar\omega_\Delta}}\,.
\end{equation}
As can be seen from the expressions above, in order to explicitly compute $\gamma^*$ we need to know on which segment $s_k(x)$ to evaluate the thermomajorization curve, i.e. we need to compute the image of $x=Z_\Delta(\beta)$ under the map $k^*$. Since this is impractical from an analytical point of view, we make use of the conditions
\begin{equation}
    \frac{\hbar \omega_0}{E_b} \ll 1\,,\quad \frac{\hbar \omega_\Delta}{E_b - \Delta} \ll 1\,,
\end{equation}
already introduced in previus sections and in the main body, where $E_b=\mathcal{E}_0(\pi/2)$ is the ground state barrier energy. In this limit, the number of segments associated to each harmonic well goes to infinity, their length goes to zero, while the total length of the curve stays finite. Thus, we can safely replace the thermomajorization curve $L_i(x)$, which is piecewise linear, with a smooth concave function $\mathcal{L}_i(x)$ allowing us to easily find an analytical expression for $\gamma^*$.
To this end, we consider the two functions
\begin{equation}
\begin{split}
      & y(t) = q + (1-q)\Big(1 - e^{-\tilde{\beta} \hbar \omega_0 t}\Big)\,,\\
      & x(t) = \frac{1-e^{-\beta \hbar \omega_0 t}}{1-e^{-\beta \hbar \omega_0}}\,,
\end{split}
\end{equation}
taking values $y_k,x_k$ when evaluated on integers $t=k$. The smooth curve $\mathcal{L}_i(x)$ is obtained by turning the parametric expression above into an explicit function $y(x)$, i.e. by eliminating the parameter $t$. Indeed, by using the inverse relation $t(x)$
\begin{equation}
   t = -\frac{1}{\beta\hbar \omega_0}\ln\left( 1- (1-e^{-\beta \hbar \omega_0}) x \right)\,,
\end{equation}
we find
\begin{equation}
\begin{split}
    y(x) = q + (1-q)\Big(1 - e^{+\frac{\tilde{\beta}}{\beta} \ln(1-(1-e^{-\beta \hbar \omega_0})x)}\Big)\,\\
   = q + (1-q) \Big( 1 - \big( 1 - (1-e^{-\beta \hbar \omega_0})x \big) ^{\tilde{\beta}/\beta} \Big) \,.
\end{split}
\end{equation}
The corresponding optimal yield can be found by setting $x+e^{-\beta W}=Z_\Delta(\beta)$:
\begin{equation}
    \gamma^* = y\left(\frac{e^{-\beta \Delta}}{1-e^{-\beta\hbar\omega_\Delta}}-e^{-\beta W}\right)\,.
\end{equation}
which leads to
\begin{equation}
    \gamma^* = q+ (1-q) \left[1-\left[1-\left(1-e^{-\beta\omega_0}\right) \left(\frac{e^{-\beta  \Delta }}{1-e^{-\beta \omega_\Delta}}-e^{-\beta W }\right)\right]^{\frac{\tilde{\beta}}{\beta} }\right]\,.
\end{equation}


\end{document}